\documentclass{svjour3}
\journalname{springer}

\begin{document}
\title{On some topological deformation of stationary spacetimes 
}


\author{ Arka Prabha Banik         \and
        Eduardo Guendelman 
}


\institute{Arka Prabha Banik \at
              Department of Physics and Astronomy, University of Missouri - Columbia , Columbia, MO, USA\\
              \email{apb3hf@mail.missouri.edu}           
           \and
            Eduardo Guendelman \at
              Department of Physics, Ben-Gurion University of the Negev, Be'er-Sheva, Israel , 84105\\
                \email{guendel@bgu.ac.il}}

\date{Received: date / Revised: date}

\maketitle

\begin{abstract}
We define a completely new space-time starting from the well known Schwarzschild Space time by defining a new polar angle $\phi '= \phi - \omega t$ and then redefining the periodicity: instead of demanding that the original angle be periodic, we demand that the new angle $\phi'$ be periodic, with period $2\pi$. This defines the "topologically rotating Schwarzchild space", which  is physically different from the standard Schwarzschild space. For this space, we work out some properties of the geodesics and related properties.This method of generating solutions can be used also for the Reissner-Nordstrom case, both in the case of Reissner-Nordstrom Black hole as well as in the case where there are no horizons, the supercharged case. Horizon shall exist in this case, but with a real singularity, not removable one by a  transformation in coordinate at the radius of the horizon of the original metric. This solution should be used as an external solution rather than the internal one. Another  topic to notice is that the improper coordinate transformation that we consider introduces closed time like curves. The noticeable topic is  that the improper coordinate transformation introduces closed time like curves which we can possibly find here too. This is a common effect in rotating spacetimes, noticeable the Godel universe and others.

\keywords{Schwarzschild Space time \and Rotating solution \and Geodesics}
\end{abstract}

\section{Introduction}
\label{intro}

It is well known that by changing boundary conditions, like the periodicity of the angle $\phi$ in spherical coordinates, we can obtain a physically different space time. In this way, for example Vilenkin \cite{Ref1} obtained the solution for a cosmic string, by starting from a flat space time 
$ds^2 = -dt^2 +r^2(d\theta^2 + sin\theta^2 d\phi^2)$
and then demanding that 
$0< \phi< 2 a \pi$
where $0<a<1$.
More generally, Stakel\cite{Ref2} has discussed this phenomenon, where space times locally defined by the same metrics could be really physically different due to different boundary conditions.Since, these two spaces are not exactly differomorphic
due to the constraints at boundary points ( there is not a one - to - one correspondence),
so we naturally expect different physics in our new spacetime.

\medskip 

As is well known the Schwarzchild metric reads as follows 
\[
ds^{2}=-(1-\frac{r_{s}}{r})dt^{2}+(1-\frac{r_{s}}{r})^{-1}dr^{2}+r^{2}(d\theta^{2}+sin^{2}\theta d\phi^{2}),
\]
with signature $+2$ and $r_{s}$ is the Schwarzchild radius.

\

Now with the transformation $\phi=\phi'+\omega t$ which also implies
$d\phi=d\phi'+\omega dt$ Schwarzchild metric changes like

\begin{equation}
ds^{2}=-(1-\frac{r_{s}}{r})dt^{2}+(1-\frac{r_{s}}{r})^{-1}dr^{2}+r^{2}(d\theta^{2}+sin^{2}\theta[d\phi'+\omega dt]^{2}),
\end{equation}

\begin{equation}
\qquad\qquad\quad=-(1-\frac{r_{s}}{r})dt^{2}+(1-\frac{r_{s}}{r})^{-1}dr^{2}+r^{2}(d\theta^{2}+sin^{2}\theta[d\phi'^{2}+2\omega d\phi dt+\omega^{2}dt^{2}]),
\end{equation}
\
finally rearranging these terms we have 

\begin{equation}
ds^{2}=-(1-\frac{r_{s}}{r}-\omega^{2}r^{2}sin^{2}\theta)dt^{2}+(1-\frac{r_{s}}{r})^{-1}dr^{2}+2\omega r^{2}sin^{2}\theta d\phi'dt+r^{2}(d\theta^{2}+sin^{2}\theta d\phi'{}^{2}),
\end{equation}
Obviously the rotated spacetime depicts a new spacetime with different
Physical and Geometrical properties than the original Schwarzchild
Spacetime.

\medskip

The new rotated space time metric has geodesics that are physically inequivalent but which still can be derived  from the previous Schwarzschild geodesics with the incorporation of the transformation $\phi' = \phi - \omega t$. The physical inequivalence derives from the fact that it is the periodicity in $\phi'$ rather than the periodicity on $\phi$ what determines when we have gone around completely in  angle $\phi'$ (rather than in the old angle $\phi$. From the new sets of equations of motions or Geodesics of this new rotating spacetime, we can investigate the Christioffel symbols. The Relativistic elliptical motion in Schwarzschild Spacetime shall become  Precessional ellipse with the addition of perturbation in $\phi$ in the form of $\omega t$. We shall calculate the change in the polar angle, i.e., $\Delta \phi'$ in a situation to make the elliptical orbit to  precess for small perturbation $\omega$.

\section{Calculation of Geodesics}

We, hereby, are  going to apply rotation $ \phi '= \phi - \omega t $ with $0< \phi'< 2 \pi$ to the already existing geodesics equations for the Schwarzschild Spacetime \cite{Ref4} and trying to find new sets of equations which shall correspond to the new topologically rotating spacetime. The four geodesics equations that comes from the Euler Lagrangian equation of motion from the Lagrangian  for  Schwarzschild Spacetime  are as follows : 
 
\begin{equation}
\frac{d^{2}t}{d\lambda^{2}}+\frac{r_{s}/r^{2}}{1-r_{s}/r}\frac{dt}{d\lambda}\frac{dr}{d\lambda}=0,
\end{equation}

\begin{equation}
\frac{d^{2}r}{d\lambda^{2}}+\frac{r_{s}}{2r^2}(1-r_{s}/r)(\frac{dt}{d\lambda})^{2}-\frac{r_{s}/r^{2}}{1-r_{s}/r}(\frac{dr}{d\lambda})^{2}-r(1-r_{s}/r)[(\frac{d\theta}{d\lambda})^{2}+sin^{2}\theta(\frac{d\phi}{d\lambda})^{2}]=0,
\end{equation}

\begin{equation}
\frac{d^{2}\theta}{d\lambda^{2}}+\frac{2}{r}(\frac{dr}{d\lambda})(\frac{d\theta}{d\lambda})-sin\theta cos\theta(\frac{d\phi}{d\lambda})^{2}=0,
\end{equation}
 
\begin{equation}
\frac{d^{2}\phi}{d\lambda^{2}}+\frac{2}{r}(\frac{dr}{d\lambda})(\frac{d\phi}{d\lambda})+2\frac{cos\theta}{sin\theta}(\frac{d\theta}{d\lambda})(\frac{d\phi}{d\lambda})=0,
\end{equation}

\
Now, incorporating the rotation in polar coordinate $ \phi '= \phi - \omega t $ or $$\ensuremath{\frac{d\phi'}{d\lambda}=\frac{d\phi}{d\lambda}-\omega\frac{dt}{d\lambda}}$$ in the previous equations, we get :

\begin{equation}
\frac{d^{2}t}{d\lambda^{2}}+\frac{r_{s}/r^{2}}{1-r_{s}/r}\frac{dt}{d\lambda}\frac{dr}{d\lambda}=0,
\end{equation}

\begin{equation}
\frac{d^{2}r}{d\lambda^{2}}+\frac{r_{s}}{2r^2}(1-r_{s}/r)(\frac{dt}{d\lambda})^{2}-\frac{r_{s}/r^{2}}{1-r_{s}/r}(\frac{dr}{d\lambda})^{2}-r(1-r_{s}/r)[(\frac{d\theta}{d\lambda})^{2}+sin^{2}\theta\{(\ensuremath{\frac{d\phi'}{d\lambda})^{2}+2\omega(\frac{d\phi'}{d\lambda^2})(\frac{dt}{d\lambda})+\omega^{2}(\frac{dt}{d\lambda}})^{2}\}]=0,
\end{equation}

\begin{equation}
\frac{d^{2}\theta}{d\lambda^{2}}+\frac{2}{r} (\frac{dr}{d\lambda})(\frac{d\theta}{d\lambda})-sin\theta cos\theta[\{(\ensuremath{\frac{d\phi'}{d\lambda})^{2}+2\omega(\frac{d\phi'}{d\lambda})(\frac{dt}{d\lambda})+\omega^{2}(\frac{dt}{d\lambda}})^{2}\}]=0,
\end{equation}

\begin{equation}
\frac{d^{2}\phi}{d\lambda^{2}}+[\frac{2}{r} (\frac{dr}{d\lambda})+2\frac{cos\theta}{sin\theta}(\frac{d\theta}{d\lambda})](\frac{d\phi'}{d\lambda}+\omega\frac{dt}{d\lambda})=0 ,
\end{equation}

\medskip

If we turn our attention to the $\theta=\pi/2$ plane, where $sin \theta =1 \;  \& \; cos \theta =0$, our geodesic equations takes simple forms :

\begin{equation}
\frac{d^{2}t}{d\lambda^{2}}+\frac{r_{s}/r^{2}}{1-r_{s}/r}\frac{dt}{d\lambda}\frac{dr}{d\lambda}=0 ,
\end{equation}

\begin{equation}\frac{d^{2}r}{d\lambda^{2}}+\frac{r_{s}}{2r^{2}}(1-r_{s}/r)(\frac{dt}{d\lambda})^{2}-\frac{r_{s}/r^{2}}{1-r_{s}/r}(\frac{dr}{d\lambda})^{2}-r(1-r_{s}/r)[(\frac{d\theta}{d\lambda})^{2}+\{(\ensuremath{\frac{d\phi'}{d\lambda})^{2}+2\omega(\frac{d\phi'}{d\lambda})(\frac{dt}{d\lambda})+\omega^{2}(\frac{dt}{d\lambda}})^{2}\}]=0,
\end{equation}

\begin{equation}
\frac{d^{2}\theta}{d\lambda^{2}}+\frac{2}{r}(\frac{dr}{d\lambda})(\frac{d\theta}{d\lambda})=\frac{d}{d\lambda}(r^{2}\frac{d\theta}{d\lambda})=0 ,
\end{equation}

\begin{equation}
\frac{d^{2}\phi}{d\lambda^{2}}+\frac{2}{r}(\frac{dr}{d\lambda})](\frac{d\phi'}{d\lambda}+\omega\frac{dt}{d\lambda})=\frac{d}{d\lambda}(r^{2}\frac{d\phi}{d\lambda})=0 .
\end{equation}

\medskip
\noindent
The last equation as can easily be seen is the Principle of Conservation of Angular Momentum.

\section{Finding Christoffel Symbols}

The standard form of Geodesic equations involves Christoffel symbols in the following manner :

\begin{equation}
\frac{d^{2}x^{\mu}}{d\lambda^{2}}+\Gamma_{\nu\alpha}^{\mu}\frac{dx^{\nu}}{d\lambda}\frac{dx^{\alpha}}{d\lambda}=0
\end{equation}
with $\mu , \nu , \alpha = 0(1)3$ and $\lambda $ is a parameter.

\medskip

So, from the Geodesic equations of motions, we get Christoffel symbols for this topologically rotating spacetime written explicitly below:

\medskip

The first Geodesic equation

\begin{equation}
\frac{d^{2}t}{d\lambda^{2}}+\frac{r_{s}/r^{2}}{1-r_{s}/r}\frac{dt}{d\lambda}\frac{dr}{d\lambda}=0,
\end{equation}

\medskip
\noindent
comparing,
 
\begin{equation}
\Gamma_{tr}^{t}=\frac{1}{2}\frac{r_{s}/r^{2}}{1-r_{s}/r} ,
\end{equation}

the Second one

\begin{equation}
\frac{d^{2}r}{d\lambda^{2}}+\frac{r_{s}}{2r^2}(1-r_{s}/r)(\frac{dt}{d\lambda})^{2}-\frac{r_{s}/r^{2}}{1-r_{s}/r}(\frac{dr}{d\lambda})^{2}-r(1-r_{s}/r)[(\frac{d\theta}{d\lambda})^{2}+sin^{2}\theta\{(\ensuremath{\frac{d\phi'}{d\lambda})^{2}+2\omega(\frac{d\phi'}{d\lambda^2})(\frac{dt}{d\lambda})+\omega^{2}(\frac{dt}{d\lambda}})^{2}\}]=0.
\end{equation}

\
By comparison, we get

\begin{equation}
\Gamma_{tt}^{r}=\frac{r_{s}}{2r^{2}}(1-\frac{r_{s}}{r})- \omega^{2} r sin^{2}\theta(1-\frac{r_{s}}{r}) ,
\end
{equation}

\begin{equation}
\Gamma_{rr}^{r}=-\frac{\frac{r_{s}}{2r^{2}}}{(1-\frac{r_{s}}{r})} ,
\end{equation}

\begin{equation}
\Gamma_{\theta\theta}^{r}=r(1-\frac{r_{s}}{r}) ,
\end{equation}

\begin{equation}
\Gamma_{\phi'\phi'}^{r}=sin^{2}\theta ,
\end{equation}

\begin{equation}
\Gamma_{\phi't}^{r}=- \omega r^{2}(1-\frac{r_{s}}{r})sin^{2}\theta ,
\end{equation}

\medskip
\noindent

from the third equation

\begin{equation}
\frac{d^{2}\theta}{d\lambda^{2}}+\frac{2}{r} (\frac{dr}{d\lambda})(\frac{d\theta}{d\lambda})-sin\theta cos\theta[\{(\ensuremath{\frac{d\phi'}{d\lambda})^{2}+2\omega(\frac{d\phi'}{d\lambda})(\frac{dt}{d\lambda})+\omega^{2}(\frac{dt}{d\lambda}})^{2}\}]=0 ,
\end{equation}

\medskip

We shall have

\begin{equation}
\Gamma_{r\theta}^{\theta}=\frac{1}{r} ,
\end{equation}

\begin{equation}
\Gamma_{\phi'\phi'}^{\theta}=-sin\theta cos\theta ,
\end{equation}

\begin{equation}
\Gamma_{tt}^{\theta}=-sin\theta cos\theta ,
\end{equation}

\begin{equation}
\Gamma_{\phi't'}^{\theta}=- sin\theta cos\theta ,
\end{equation}

\medskip
\noindent

The last one 
\begin{equation}
\frac{d^{2}\phi}{d\lambda^{2}}+[\frac{2}{r} (\frac{dr}{d\lambda})+2\frac{cos\theta}{sin\theta}(\frac{d\theta}{d\lambda})](\frac{d\phi'}{d\lambda}+\omega\frac{dt}{d\lambda})=0 ,
\end{equation}

\medskip

leads us to 

\begin{equation}
\Gamma_{rt}^{\phi'}=\frac{1}{2}\frac{\omega}{r}(2-\frac{r_{s}/r}{1-r_{s}/r}) ,
\end{equation}

\begin{equation}
\Gamma_{r\phi'}^{\phi}=\frac{1}{r} ,
\end{equation}

\begin{equation}
\Gamma_{\theta\phi'}^{\phi'}=\frac{ cos\theta}{sin\theta} ,
\end{equation}

\begin{equation}
\Gamma_{\theta t}^{\phi'}=\frac{ \omega cos\theta}{sin\theta}.
\end{equation}

\
These are 14 Christoffel symbols retrieved from Geodesics equations of motion of the newly formed rotated spacetime.

\section{Calculation of Precession of Perihelium}

In this section, we are going to compute the deflection of a relativistic particle in this new spacetime\cite{Ref5}. We have already got the metric for the topologically rotating Schwarzchild spacetime from the Schwarzchild spacetime itself, by a rotation in polar coordinate by $\phi '= \phi - \omega t$, reads as follows :

\begin{equation}
ds^{2}=-(1-\frac{r_{s}}{r}-\omega^{2}r^{2}sin^{2}\theta)dt^{2}+(1-\frac{r_{s}}{r})^{-1}dr^{2}+2\omega r^{2}sin^{2}\theta d\phi'dt+r^{2}(d\theta^{2}+sin^{2}\theta d\phi'{}^{2}),
\end{equation}

\noindent
So that the metric components are:

\begin{equation}
g_{tt}=-(1-\frac{r_{s}}{r}-\omega^{2}r^{2}sin^{2}\theta),
\end{equation}

\begin{equation}
g_{rr}=(1-\frac{r_{s}}{r})^{-1} ,
\end{equation}

\begin{equation}
g_{t\phi'}=\omega r^{2}sin^{2}\theta ,
\end{equation}

\begin{equation}
g_{\theta\theta}=r^{2} ,
\end{equation}

\begin{equation}
g_{\phi'\phi'}=r^{2}sin^{2}\theta ,
\end{equation}

\
We know that the Lagrangian is of the form 

\begin{equation}
L=g_{\mu\nu}\frac{dx^{\mu}}{d\tau}\frac{dx^{\nu}}{d\tau} ,
\end{equation}

\noindent
$\tau$ being parameter.

\medskip
\noindent
So, that 
\begin{equation}
L=-(1-\frac{r_{s}}{r}-\omega^{2}r^{2}sin^{2}\theta)\dot{t}^{2}+(1-\frac{r_{s}}{r})^{-1}\dot{r}^{2}+2\omega r^{2}sin^{2}\theta\dot{\phi'}\dot{t}+r^{2}\dot{\theta}^{2}+r^{2}sin^{2}\theta\dot{\phi'}{}^{2} ,
\end{equation}

\noindent
The Euler-Lagrange equation for $\theta $ will be :

\[
\frac{\partial L}{\partial\theta}-\frac{\partial}{\partial\tau}(\frac{\partial L}{\partial\theta/\partial\tau})=0 ,
\]

so

\begin{equation}
  r^{2}\ddot{\theta}+2r\dot{r}\dot{\theta}-r^{2}sin\theta cos\theta\dot{\phi}'^{2}-2\omega r^{2}sin\theta cos\theta\dot{\phi}'\dot{t}=0 ,
\end{equation}

Now, in $\theta =\pi/2$ plane where $sin\theta =1 \:\&\: cos\theta =0 $, we have the Lagrangian of the form 

$$L=-(1-\frac{r_{s}}{r}-\omega^{2}r^{2} \theta)\dot{t}^{2}+(1-\frac{r_{s}}{r})^{-1}\dot{r}^{2}+2\omega r^{2}\dot{\phi'}\dot{t}+r^{2}\dot{\theta}^{2}+r^{2}\dot{\phi'}{}^{2}$$ .

From this equation, we can see that Lagrangian doesn't involve $\phi' $ and $t$, hence $\phi' $ and $t$ are the ignorable coordinates. So, we can easily write down Euler - Lagrange equation for $\phi' $ and $t$.  
For $\phi' $ :

\[
(\frac{\partial L}{\partial \dot{\phi'}})=h ,
\]

$h$ being a constant.

Now,

\begin{equation}
  r^{2}(\dot{\phi}'+\omega\dot{t})=h ,
\end{equation}

Again, for $t$ :

\[
(\frac{\partial L}{\partial \dot{t}})=k ,
\]

\
$k$ being a constant.

Now,

\begin{equation}
-(1-\frac{r_{s}}{r}-\omega^{2}r^{2})\dot{t}+2\omega r^{2}\dot{\dot{\phi}'=K} ,
\end{equation}

\begin{equation}
 -(1-\frac{r_{s}}{r})\dot{t}=K-\omega^{2}r^{2}(\dot{\phi}'+\omega\dot{t}) ,
\end{equation}

\begin{equation}
 -(1-\frac{r_{s}}{r})\dot{t}=K-\omega^{2}h ,
\end{equation}

\begin{equation} 
 -(1-\frac{r_{s}}{r})\dot{t}=l ,
\end{equation}

where, $l=k-\omega^2 h$ is another constant.

\medskip
\noindent
Dividing the metric equation by $d\tau$  and working out at the plane $\theta=\pi/2$ with the immediate constraint of $\dot{\theta}=0=\ddot{\theta}$, we get :

\begin{equation}
-(1-\frac{r_{s}}{r}-\omega^{2}r^{2})\dot{t}^{2}+(1-\frac{r_{s}}{r})^{-1}\dot{r}^{2}+2\omega r^{2}\dot{\phi'}\dot{t}+r^{2}\dot{\phi'}{}^{2}=1 .
\end{equation}

\
By using equation (45) and (49) we have

\begin{equation}
-(1-\frac{r_{s}}{r})\dot{t}^{2}+\omega r^{2}(1-\frac{r_{s}}{r}) [\dot{\phi'}+\omega t]^{2}=(1-\frac{r_{s}}{r})-\dot{r}^{2} ,
\end{equation}

\noindent
modifying this equation we get 

\begin{equation}
\dot{r}^{2}=l^{2}+(1-\frac{h^2}{r^2})(1-\frac{r_{s}}{r}) ,
\end{equation}

\
putting $u=\frac{1}{r}$ ,

\medskip

we shall have, $\dot{r}=-\frac{1}{u^{2}}\dot{u}$ , 

\medskip

hence the equation will read as,

\begin{equation}
\frac{d^{2}u}{d\phi^{2}}+u=\frac{r_{s}}{2h^{2}}+\frac{3}{2}r_{s}u^{2} ,
\end{equation}

\
Which is an equation of relativistic elliptical orbit.

\medskip
\noindent
Replacing $u$ with the classical dominating term $u_{0}=\frac{r_{s}}{2h^{2}}(1+ecos\phi)$

\medskip
\noindent
where, $e$ is the eccentricity of the ellipse with $0<e<1$.

\medskip
\noindent
So, the differential equation will be looking like after putting $u=u_{0}$
i.e, the dominating term on the right 

\begin{equation}
\frac{d^{2}u}{d\phi^{2}}+u=\frac{r_{s}}{2h^{2}}+\frac{3r_{s}^{3}}{8h^{4}}(1+2ecos\phi+e^{2}cos^{2}\phi),
\end{equation}

\medskip
\noindent
It is to be noted minutely that the first and the last term inside
the parenthesis is only adding small perturbation or oscillation ,
so neglecting this small terms it reads

\begin{equation}
\frac{d^{2}u}{d\phi^{2}}+u=\frac{r_{s}}{2h^{2}}+\frac{3r_{s}^{3}}{4h^{4}}cos\phi ,
\end{equation}

\medskip
\noindent
The solution of this differential equation is 

\begin{equation}
u=u_{0}+\frac{3r_{s}^{3}}{2h^{4}}e\phi sin\phi ,
\end{equation}

\begin{equation}
 u=\frac{r_{s}}{2h^{2}}[1+ecos\phi+\frac{3r_{s}^{2}}{2h^{2}}e\phi sin\phi] ,
\end{equation}

\medskip
\noindent
since, $\frac{3r_{s}^{2}}{2h^{2}}\phi$ being small, with the appropriate
approximation of this term in the arguments of  $sine$ and $cosine$
, we get in our hand finally

\begin{equation}
u=\frac{r_{s}}{2h^{2}}[1+ecos(\phi-\frac{3r_{s}^{2}}{2h^{2}}\phi)],
\end{equation}

\medskip

\medskip
\noindent
This gives a precessing ellipse.

\

Now, the rotation yields us $\phi'=\phi-\omega t$

\begin{equation}
  u=\frac{r_{s}}{2h^{2}}[1+ecos(\{\phi'+\omega t\}-\frac{3r_{s}^{2}}{2h^{2}}\{\phi'+\omega t\})].
\end{equation}

\medskip
\noindent
The argument of $cosine$ changes by $2\pi$ to ellipse to precess in
the orbit.

\medskip
\noindent
So,

\begin{equation}
\Delta\phi'(1-\frac{3r_{s}^{2}}{2h^{2}})+\omega\Delta t=2\pi
\end{equation}

\medskip
\noindent
To find $\omega\Delta t(\phi')$ we start with the conservation of
angular momentum, i.e.,

\[
r^{2}\dot{\phi}=h ,
\]

so,
\begin{equation}
\dot{\phi}=hu^{2}=\frac{r_{s}^{4}}{4h^{3}}(1+2ecos\phi+e^{2}cos^{2}\phi).
\end{equation}

\medskip
\noindent
The last term is negligible since $e<<1\;\&\:\frac{r_{s}^{4}}{4h^{3}}<<1$

\begin{equation}
d\phi=cdt(1+2ecos\phi),
\end{equation}
 
so,
\begin{equation}
 \frac{d\phi}{(1+2ecos\phi)}=cdt.
\end{equation}

Where, $c=\frac{r_{s}^{4}}{4h^{3}}$

Now, again for small $e<<1\:\&\:cos\phi\leq1$

\begin{equation}
ct(\phi)=\int(1-2ecos\phi)d\phi=\phi-2sin\phi
\end{equation}

\medskip
\noindent
Now, in rotated coordinate $\phi'$

\begin{equation}
ct(\phi')=(\phi'+\omega t)-2esin(\phi'+\omega t)
\end{equation}

\medskip
\noindent
For small $\omega$, we can do Taylor expansion of the last term to
get 

\begin{equation}
ct(\phi')=(\phi'+\omega t)-2sin\phi'-2ecos\phi'\omega t
\end{equation}

For small $e$and $\omega$the last term can be neglected 

\begin{equation}
(c-\omega)t(\phi')=\phi'-2sin\phi'
\end{equation}

\begin{equation}
(c-\omega)\Delta t(\phi')=\Delta\phi'-2\Delta(sin\phi')
\end{equation}

\medskip
\noindent
Now,$\Delta\phi'$runs from $0$to $2\pi$ , but $\Delta(sin\phi')$
is small compared to that, hence neglecting this term we have 

\begin{equation}
\omega\Delta t(\phi')=\frac{1}{c/\omega-1}\Delta\phi'=\frac{1}{\frac{r_{s}^{4}}{4\omega h^{3}}-1}\Delta\phi'
\end{equation}

Putting this in equation (60), we have 

\begin{equation}
\Delta\phi'(1-\frac{3r_{s}^{2}}{2h^{2}})+\frac{1}{\frac{r_{s}^{4}}{4\omega h^{3}}-1}\Delta\phi'=2\pi ,
\end{equation}

\begin{equation}
 \Delta\phi'=\frac{2\pi}{(1-\frac{3r_{s}^{2}}{2h^{2}})+\frac{1}{\frac{r_{s}^{4}}{4\omega h^{3}}-1}} ,
\end{equation}

\medskip

Finally, we have got the expression for change in rotated polar coordinate
in the precissional elliptical orbit. A Quickcheck: in the limit $\omega\rightarrow 0$ equation (71) produces the change of rotated polar coordinate for Schwarzchild spacetime's precission.

\section{Motion of a standard clock along geodesic and discussion on Closed timelike curves (CTC)}

If we think of a moving clock along the geodesic equation of our spacetime, we can find the time difference between the two beams moving along in two opposite directions\cite{Ref11}. To find this time difference, we start with the metric
$$ ds^{2}=-(1-\frac{r_{s}}{r}-\omega^{2}r^{2}sin^{2}\theta)dt^{2}+(1-\frac{r_{s}}{r})^{-1}dr^{2}+2\omega r^{2}sin^{2}\theta d\phi'dt+r^{2}(d\theta^{2}+sin^{2}\theta d\phi'{}^{2}) , $$
\
The Keplarian period of the orbit is $T_0=\frac{2\pi }{\omega_0}$, with $\omega_0=\sqrt{\frac{M}{r^3}}$

For fixed $r$ and $\theta =\pi/2$ , we have the geodesic equation:
\begin{equation}
\frac{M}{r^{2}}(1-\frac{2M}{r})dt^{2}-r(1-\frac{2M}{r})+[d\phi'+\omega dt]^{2}=0 ,
\end{equation}

where we have considered it with $r_{s}=2M$.

This yields

\begin{equation}
d\phi'+\omega dt=\pm\sqrt{\frac{M}{r^{3}}}dt ,
\end{equation}
 
 which yields

\begin{equation}
\frac{dt}{d\phi'}=\frac{1}{-\omega\pm\omega_{0}}.
\end{equation}

So that 

\begin{equation}
t_{\pm}=\frac{2\pi}{-\omega\pm\omega_{0}}=\frac{T_{0}}{-\frac{\omega}{\omega_{0}}\pm1} ,
\end{equation}

Now, its easy to calculate the time difference which is 

\begin{equation}
\Delta t=t_{+}-t_{-}=2T_{0}[\frac{1}{1-(\frac{\omega}{\omega_{0}})^{2}}].
\end{equation}

In the limit of small $\omega$ it makes 

$$
\Delta t=2T_{0}.
$$
\
\medskip

Next, We can see some behaviour times for different values of $\omega$ and
$\omega_{0}$.
\begin{itemize}
\item For $\omega=\omega_{0}$, $t_{\pm}\rightarrow\infty$ 
\item For $\mid\frac{\omega}{\omega_{0}}\mid>1$, $t_{+}$and $t_{-}$ have
different signs.
\item For $\mid\frac{\omega}{\omega_{0}}\mid<1$, $t_{+}$and $t_{-}$ have
same signs.
\end{itemize}

\
\medskip

From the part of time delay along geodesics, we can understand that $t_{+},t_{-}$ are the times for two opposite directions.

\medskip

Now, suppose, $t_{+}$ and $t_{-}$ is for the clockwise and anticlockwise
respectively.
For $\omega=\omega_{0}$, one of these will go to infinity. More specifically,
For $\omega>0$, $t_{+}$ will go to infinity.

\medskip
To show the existence of CTC, we start by observing  that, as $r\rightarrow\infty$,
$\omega_{0}\rightarrow0$. In this condition, 
\begin{equation}
t_{\pm}=-\frac{2\pi}{\omega}.
\end{equation}

\medskip
 This equation corresponds to the geodesics which in original coordinates
had $\phi=constant=0$( say ).
Since we know in the new coordinate $\phi'=\phi-\omega t$. So that $$t=-\frac{\phi'}{\omega}$$. Since $\phi'$ is periodic so it makes $t$ to be periodic too. This means the existence of closed timelike
curve.

\
\medskip
Notice that this CTC appears because the particle is not rotating fast enough. If the radius is small enough the geodesics can become " normal" timelike curves.

\
\medskip

Suppose, if the orbit manages to go around in the angular variable fast enough, that is
the frequency of the rotating spacetime is not more than the  Keplerian frequency $\omega_{0}$ ,then, trajectory is like a 4d spiral in space time which is not a closed time like curve. In the language of mathematics, we can then determine the critical radius by the following condition
\

$$\omega \leq \omega_{0}.$$

Therefore, the normal time like curve may lie below 

\
\noindent

Since, $\omega_{0}=\frac{\sqrt{M}}{\sqrt{r^{3}}} ,$  

\
\noindent

hence,  $$r\leq  \sqrt[3]{\frac{M}{\omega^{2}}}.$$

\medskip

Note all the calculations has been applicable for the distance greater than the critical radius.

\section{Calculation of the Sagnac time delay }

This section is dedicated to the calculation of the Sagnac time delay for this new rotating  Schwarzschild Space time. We shall here follow Culetu again \cite{Ref2}. Sagnac time is to be calculated by assumption that a source of light is fixed and two oppositely directed beams produced from a mirror  interfereto yield shift in angular velocity in these two beams and hence the time delay gets its introduction.

We have our metric 
$$ ds^{2}=-(1-\frac{r_{s}}{r}-\omega^{2}r^{2}sin^{2}\theta)dt^{2}+(1-\frac{r_{s}}{r})^{-1}dr^{2}+2\omega r^{2}sin^{2}\theta d\phi'dt+r^{2}(d\theta^{2}+sin^{2}\theta d\phi'{}^{2}) .$$
\
By choosing the source is at fixed  $r=r_0$  and $\theta = \theta_0$, then dividing the metric with $dt^2$ and followed by  equating with  $0$ , we shall have 
\begin{equation}(\frac{d\phi'}{dt})^{2}+2\omega(\frac{d\phi'}{dt})-\frac{1}{r_{0}^{2}sin^{2}\theta_{0}}(1-\frac{r_{s}}{r_{0}}-\omega^{2}r_{0}^{2}sin^{2}\theta_{0})=0 .
\end{equation}

which is a quadratic equation.

\
If $\Omega_{\pm}=(\frac{d\phi'}{dt})_{\pm}$ are the roots of this equation.

\
then $\Omega_{\pm}=-\omega\pm\sqrt{\omega^{2}+\frac{1}{r_{0}^{2}sin^{2}\theta_{0}}(1-\frac{r_{s}}{r_{0}}-\omega^{2}r_{0}^{2}sin^{2}\theta_{0})},$

so,

$\Omega_{\pm}=-\omega\pm\sqrt{\frac{1}{r_{0}^{2}sin^{2}\theta_{0}}(1-\frac{r_{s}}{r_{0}})}$

hence,

$ \Omega_{\pm}=-\omega\pm\frac{1}{r_{0}sin\theta_{0}}(1-\frac{r_{s}}{r_{0}})^{1/2}$

\
\noindent

This gives the Sagnac time delay 

\begin{equation}
\Delta t=t_{+}-t_{-}=2\pi(\Omega_{+}^{-1}-\Omega_{-}^{-1}),
\end{equation}

\begin{equation}
=4\pi(\frac{\frac{1}{r_{0}sin\theta_{0}}(1-\frac{r_{s}}{r_{0}})^{1/2}}{\frac{1}{r_{0}^{2}sin^{2}\theta_{0}}(1-\frac{r_{s}}{r_{0}})-\omega^{2}}),
\end{equation}
Note,this time delay is dependent upon $\omega$.

\section{Mixing of t and $\phi$ and Kantowsky - Sachs Spacetime}

In a Schwarzchild desitter spacetime, where $g_{00}=-(1-2\frac{GM}{r}-\Lambda\frac{r^{2}}{3})$ and $g_{rr}=(1-2\frac{GM}{r}-\Lambda\frac{r^{2}}{3})^{-1}$ , for sufficiently large $ \Lambda$ and $M$ or inside the horizon $t$ and $\phi$ are both space like dimension in all space time. As, for $\Lambda$  and $M$ being big enough, then $1-(2 \frac{GM}{r} )-(\Lambda \frac{r^2}{3} )$ is negative everywhere in the spacetime, which is called Kantowsky-Sachs spacetime.
 
\
\noindent

So, that $t$ is a spatial coordinate (and of course $\phi$ also is a spatial coordinate) then everywhere in spacetime  and by making linear combinations and redefining periodicities,  we are not at risk of producing closed time like curves. Therefore, we can consider that inside the horizon this transformation doesn't introduce closed timelike curves. In concluding remark, we can add that we have internal solution for this transformed spacetime where its highly unlikely to have closed timelike curves.

\section{Conclusions}

We have defined a new space-time starting from the well known Schwarzschild Space time by defining a new polar angle $\phi ' =\phi - \omega t$ and then redefining the periodicity: instead of demanding that the original angle be periodic, we demand that the new angle $\phi'$ be periodic, with period $2\pi$. This defines the "topologically rotating Schwarzchild space", which  is physically different from the standard Schwarzschild space. for this space we work out some properties of the geodesics and related properties. This method of generating solutions can be used also for the Reissner-Nordstrom case, both in the case of Reissner-Nordstrom Black hole as well as in the case there are no horizons, the supercharged case. 

\medskip

In this case, there are horizon, the  transformation  $\phi ' =\phi - \omega t$  involves the  time $t$ which diverges at the horizon, therefore, while at this point the angle $\phi'$ has periodicity $2 \pi$, the period of the old angle $\phi$ diverges.  The procedure introduces a real singularity, not removable any more by a coordinate transformation at the radius of the horizon of the original metric, introducing Kruskal coordinates, for example, does not make the metric smooth at the horizon, therefore, the new solution is not a Black Hole.

\medskip
The physical meaning of this singularity is the presence of light like matter at the Schwarzschild radius.
\
There in lies a rotating solution, where away from  some surface we use the rotating Schwarzschild solution,but also inside there is another solution. In this case the problem of the t coordinate being bad at $r=2GM$ is not a problem, since we are using the the rotating Schwarzschild solution as an external solution, never applying it to regions where $r$ of the order of $2GM$.

\medskip

Here, we have shown Sagnac time delay along with the time delay of a moving clock  along a closed geodesics. The Precession of Perhilium in this new spacetime has been calculated precisely.



\end{document}